\documentclass{iau} 
\usepackage{graphicx}
\usepackage{float}
\usepackage[flushleft]{threeparttable}

\title[Using planetary transits to estimate magnetic cycles lengths in Kepler stars] 
{Using planetary transits to estimate magnetic cycles lengths in Kepler stars}

\author[Raissa Estrela \& Adriana Valio]   
{Raissa Estrela$^1$
 \and Adriana Valio$^1$}

\affiliation{$^1$Center for Radio Astronomy and Astrophysics (CRAAM), Mackenzie Presbyterian University, Sao Paulo, Brazil \\  Rua da Consolacao 01301-000, 896, Sao Paulo, Brazil \\ email: {\tt rlf.estrela@gmail.com} \\[\affilskip]
email: {\tt avalio@craam.mackenzie.br}}

\pubyear{2016}
\volume{328}  
\jname{Living around active stars}
\editors{D. Nandi, A. Valio \& P. Petit, eds}
\begin{document}

\maketitle

\begin{abstract}
Observations of various solar-type stars along decades revealed that they can have magnetic cycles, just like our Sun. An investigation of the relation between their cycle length and rotation period can shed light on the dynamo mechanisms operating in these stars. Previous works on this relation suggested that the stars could be separated into active and inactive branches, with the Sun falling between them. In this work, we determined short magnetic activity cycles for 6 active solar-type stars observed by the Kepler telescope. The method adopted here estimates the activity from the excess in the residuals of the transitlight curves. This excess is obtained by subtracting a spotless model transit from the light curve, and then integrating over all the residuals during the transit. The presence of long term periodicity is estimated from the analysis of a Lomb-Scargle periodogram of the complete time series. Finally, we investigate the rotation-cycle period relation for the stars analysed here and find that some active stars do not follow the behaviour proposed earlier, falling in the inactive branch. In addition, we also notice a considerable spread from other stars in the literature in the active/inactive branches.

\end{abstract}



\firstsection 
\section{Introduction}

The magnetic activity of the Sun varies throughout the 22 year long magnetic cycle, with the polarity of its magnetic field flipping every 11 years. These cycles are identified by the frequency and number of sunspots in the solar surface, which acts as an indicator of activity. Stellar activity is also present in other stars, that show remarkable lightcurve variations due to starspots and other magnetic phenomena. \cite[Skumanich (1972)]{Skumanich72} first suggested that the activity of the star was associated with its rotation rate, and consequently with its age. Therefore, young rapidly rotating stars show higher level of activity and can produce larger spots and energetic flares.

The Mount Wilson Observatory Ca II H K survey was the first to show that hundred of stars could also exhibit long and short periodic cycles between 2.5 and 25 years \cite[(Baliunas et al., 1995)]{bau95}. Using this data, \cite[Saar \& Brandenburg (1999)]{saar99} established a relation between the stellar rotation period, P$_{\rm orb}$ and stellar cycle period, P$_{\rm cycle}$ as a function of the Rossby number. This relation divided the stars in two branches: active (A) and inactive (I), according to its activity level and rotation rate. The active sequence is composed by stars that rotates faster than the Sun, while the inactive one has slow rotating stars. Later, it was observed that some of the stars in the active branch also exhibits secondary short cycles that fall in the inactive branch \cite[(B{\"o}hm-Vitense, 2007)]{bohm07}.

Recently, the Kepler telescope provided long-term high photometric precision of thousand of stars. These data offers an unique opportunity to increase our understanding about magnetic cycles in other stars. In addition, it could improve stellar dynamo models in different type of stars. Among the studies of magnetic cycles using Kepler data, \cite[Vida \& Olah (2014)]{Vida2013} analyzed fast-rotating stars and found activity cycles of 300-900 days for 9 targets, and \cite[Marthur et al. (2014)]{Marthur_etal14} found evidences of magnetic cycle in two Kepler solar type star.

\section{Spot model}

Kepler-17 and Kepler-63 are two active solar-type stars that exhibit rotational modulations in their light curve caused by the presence of starspots, with a peak-to-peak variation of $6\%$ and $4\%$, respectively.
To analyze and characterize the physical parameters of the spots in these stars, we applied the transit model proposed by Silva (2003).
This model simulates the passage of a planet in front of its host star. The simulation consider a star with quadratic limb darkening as a 2D image and the planet is assumed to be a dark disk with radius $R_{p}/R_{\rm star}$, where R$_p$ is the radius of the planet and $R_{\rm star}$ is the radius of the primary star. The sum of all the pixels in the image (star plus dark planet) yields the transit light curve. 

We added round spots to the stellar surface and modelled them by three parameters: intensity (in function of the intensity at disc center I$_{c}$), radius (measured in units of planetary radius) and position (longitude and latitude). The latitude remains fixed and equal to the transit latitude. The longitude of the spots is limited to -70 and 70$\,^{\circ}$ from the central meridian to avoid any distortions caused by the ingress and the egress of the transit. In the case of Kepler-17, the latitude is -14$\,^{\circ}$.6, while for Kepler-63b the planet occults several latitudes of the star from its equator all the way to the poles due to its high obliquity. 

To obtain a better fit for each transit light curve, it was necessary to refine the values for the semi-major axis and planet radius obtained from literature for Kepler-17b \cite[(D{\'e}sert et al., 2011)]{desert11} and for Kepler-63b \cite[(Sanchis-Ojeda et al., 2013)]{sanchis13}. The physical parameters of the stars are described in Table 1.

\begin{figure*}[ht]
  \centering
\includegraphics[scale=0.40]{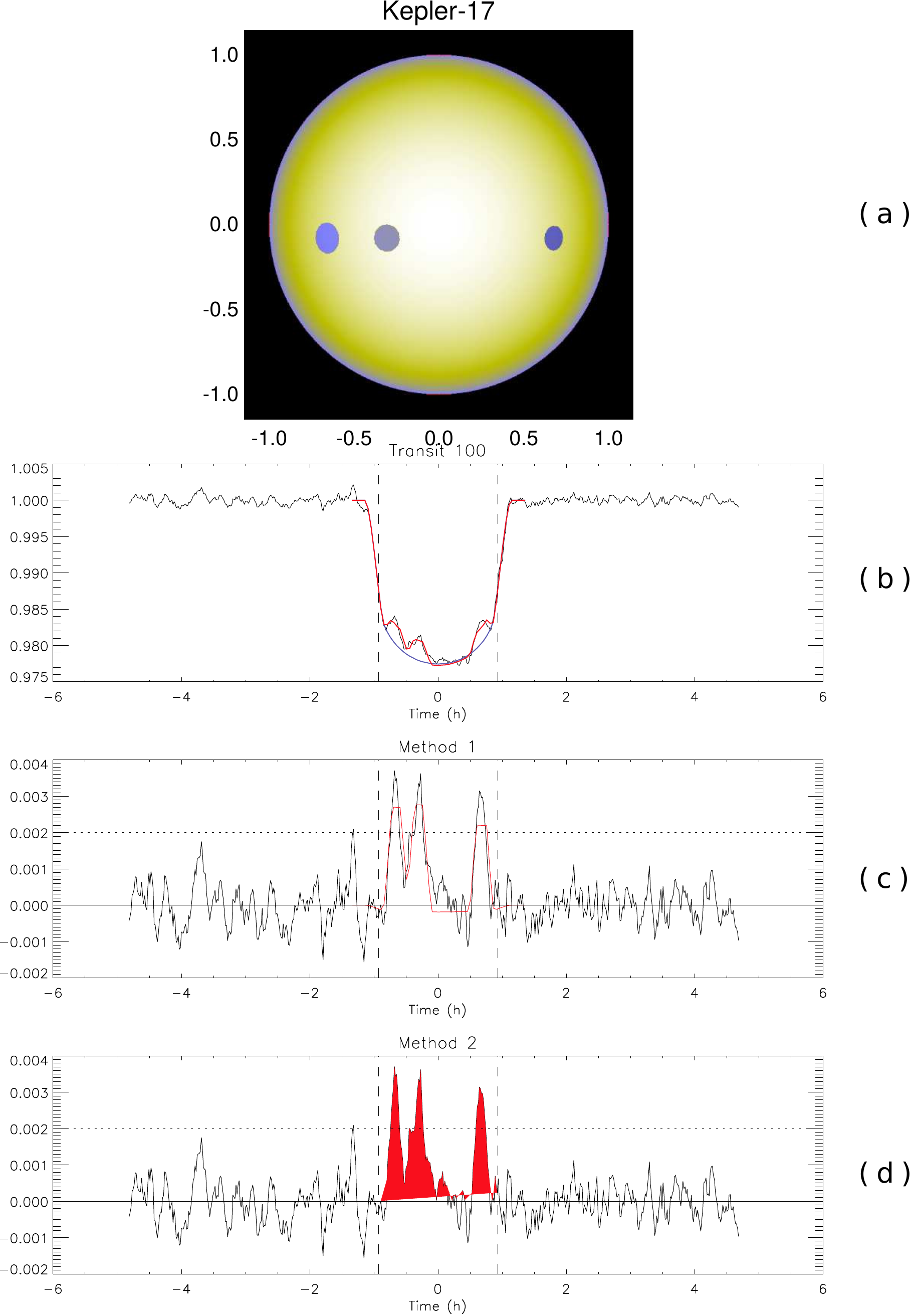}
  \caption[Model]{The 100th transit from Kepler-17 illustrates a typical example of the two methods adopted in this work: the spot model developed by \cite[Silva (2003)]{sil03}  and transit residual excess. (a): Synthetised star with three spots. (b): Transit light curve with the model of a spotless star overplotted (red). (c) Residuals of the transit lightcurve after subtraction of a spotless star model. The red curve shows the fit to the data ``bumps''. (d): Integration (in red) of the residual excess resulted from the subtraction. } 
\label{fig:figura1_1}
\end{figure*}

To fit the spots, we subtracted a spotless model from the transit light curve. The result of this subtraction are the residuals, where the spots became more evident, as the ``bumps'' seen in the residuals. This process is illustrated in Fig. \ref{fig:figura1_1}c for Kepler-17. We used the CDPP (Combined Differential Photometric Precision, see \cite[Christiansen et al. (2012)]{christiansen12} as an estimation of the noise in the Kepler data. Each quarter of the light curve has an associated CDPP value. Here, we considered the uncertainty in the data, $\sigma$, as being ten times the average of the CDPP values in all quarters. Only the ``bumps'' that exceed the detection limit of 10$\sigma$ are assumed as spots and modelled. Finally, the best fit of the spots parameters is obtained by minimizing $\chi^2$, calculated using the AMOEBA routine \cite[(Press et al., 1992)]{press92}.

\begin{table*}[!h]
\renewcommand{\arraystretch}{2}
\label{table:tab1}
\resizebox{0.9\textwidth}{!}{\begin{minipage}{\textwidth}
\centering
\caption{}
\begin{tabular}{lccccl}
\multicolumn{6}{c}{\textbf{Observational parameters of the stars}}                                                                                                                                                       \\
\hline  
                    & \multicolumn{1}{l}{Radius {[}R$_\odot${]}} & \multicolumn{1}{l}{Age {[}Gyr{]}} & \multicolumn{1}{l}{Effective  Temperature {[}K{]}} & \multicolumn{1}{l}{Rotation Period {[}days{]}} & Reference \\                  
Kepler-17           & 1.05 $\pm$ 0.03                              & $<$ 1.78                          & 5780 $\pm$ 80                                      & 11.89                                          & 1,2       \\
Kepler-63           & 0.901$^{+ 0.022}_{-0.027}$                   & 0.2                               & 5580 $\pm$ 50                                      & 5.40                                           & 3         \\
KIC 9705459         & 0.951 $^{+0.159}_{-0.04}$                    &                                   & 5900$^{+106}_{-125}$                               & 2.83                                           & 4,5       \\
KIC 5376836         & 0.885 $^{+0.363}_{-0.081}$                   &                                   & 5903$^{+93}_{-112}$                                & $\sim$ 4                                       & 4,5       \\
Kepler-96           & 1.02 $\pm$ 0.09                              & 2.34                              & 5690.0 $\pm$ 73.0                                   & 15.30                                          & 5,6       \\
Hat-p-11 (Kepler-3) & 0.75 $\pm$ 0.02                               & 6.5$^{+4.1}_{-4.1}$               & 4780.0 $\pm$ 50.0                                  & 30                                             & 7        
\end{tabular}
\begin{threeparttable}
\begin{tablenotes}
 \item[1] \textbf{References.} (1) \cite[Bonomo et al. (2012)]{bon12}, (2) \cite[D{\'e}sert et al. (2011)]{desert11}, (3) \cite[Sanchis-Ojeda et al. (2013)]{sanchis13}, (4) MAST Kepler database, (5) \cite[Walkowicz \& Basri (2013)]{wal13}, (6) \cite[Marcy et al. (2014)]{marcy14} and (7) \cite[Sanchis-Ojeda \& Winn (2011)]{sanchis11}.
\end{tablenotes}
\end{threeparttable}
\end{minipage}}
\end{table*}

To determine the stellar cycles we used two approaches \cite[(Estrela \& Valio (2016))]{estrela16}. The first one is the analysis of the variation in the number of spots during the 4 years of observation of the Kepler stars. The latter is the calculation of the flux deficit resulting from the presence of spots on the star surface. The relative flux deficit of a single spot is the product of the spot contrast and its area, thus for each transit the total flux deficit associated with spots was calculated by summing all individuals spots: 

\begin{equation}
F \approx \sum (1-f_{i})(R_{\rm spot})^{2}
\end{equation} 

where the spots contrast is taken to be (1-f$_{i}$), and f$_{i}$ is the relative intensity of the spot with respect to the disk center intensity I$_{c}$. A value of f$_{i}$ = 1 means that there is no spot at all. 

Possible long duration trends were removed in these time series by applying a quadratic polynomial fit and then subtracting it. Then, a Lomb Scargle periodogram (LS) \cite[(Scargle, 1982)]{scargle82} was applied on these time series to obtain the period related to the magnetic cycle. In addition, it was applied a significance test to quantify the significance of the peaks from the LS periodogram. The statistical significance associated to each frequency in the periodogram is determined by the p-value (\textrm{p}). The smaller the p-value, the larger the significance of the peak. We adopted the significance level $\alpha$ as being 3$\sigma$ (\textrm{p} $\pm$ 0.0013). Each periodogram in Figure \ref{fig:figura1_2} has a significance test associated (plotted below). The uncertainty of the peaks in the periodogram is given by the FWHM of the peak power.

\begin{figure*}[!h]
  \centering
\includegraphics[scale=0.40]{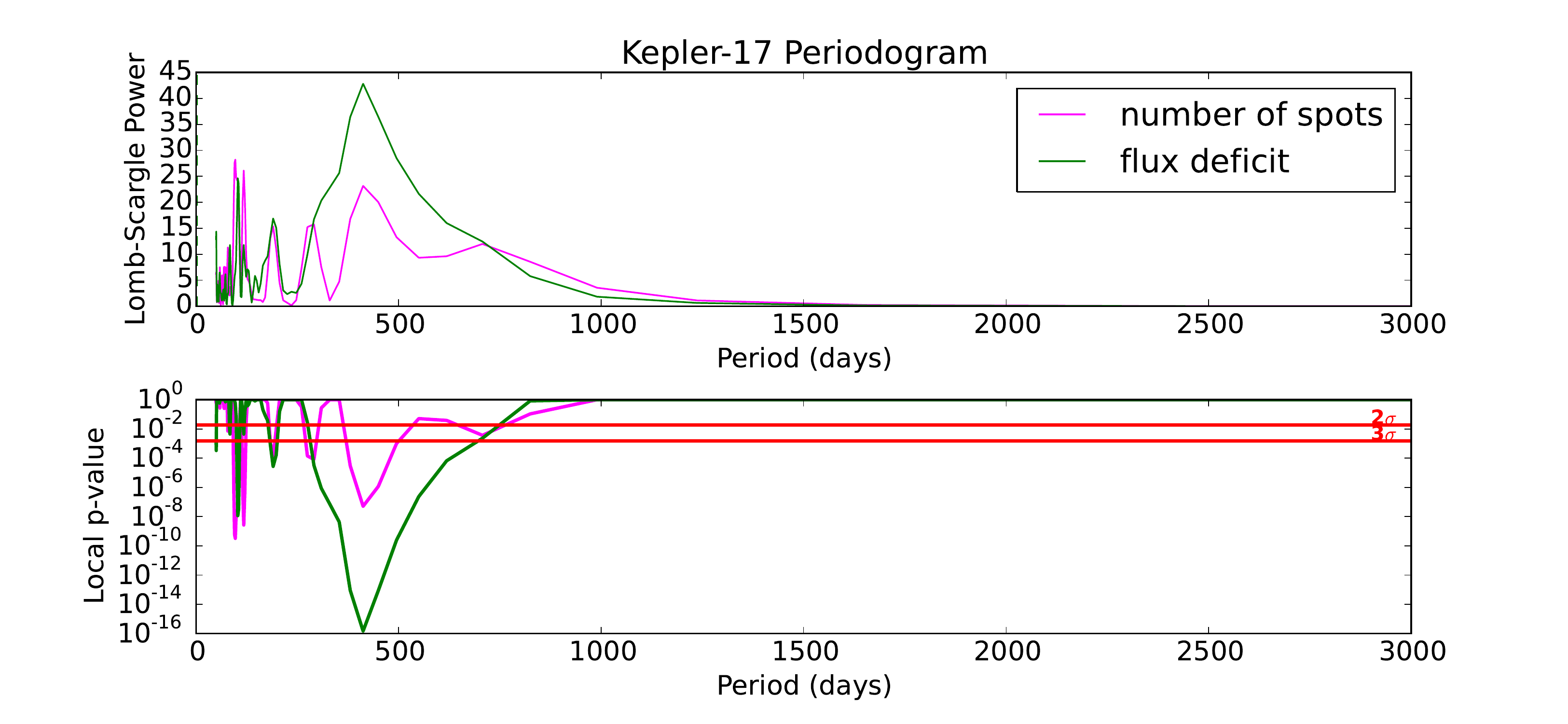}
\includegraphics[scale=0.40]{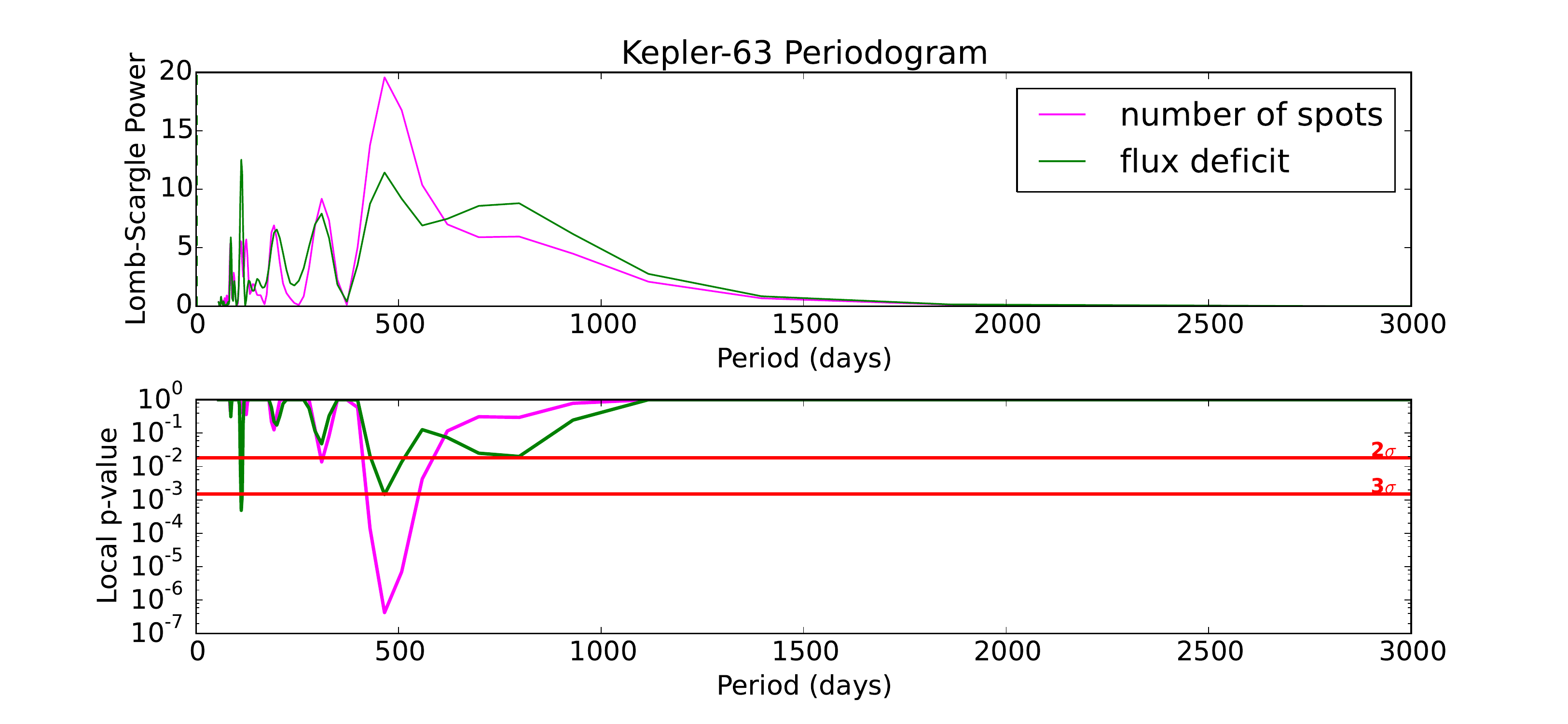}
  \caption[Lomb]{Lomb Scargle periodogram applied to the number of spots and total flux deficit of Kepler-17 (\textit{top}) and Kepler-63 (\textit{bottom}). The highest peak, indicated by a dashed line, corresponds to a periodicity of 410 $\pm$ 50 days for Kepler-17 and 460 $\pm$ 60 days for Kepler-63.}
\label{fig:figura1_2}
\end{figure*}

The LS periodogram detected a long term periodicity for both stars, as shown in Fig.~\ref{fig:figura1_2}. Kepler-17 shows a prominent peak at 410 $\pm$ 60 days (number of spots) and 410 $\pm$ 50 days (flux deficit), while Kepler-63 shows a periodicity of 460 $\pm$ 60 days for the total number of spots, and 460 $\pm$ 50 days for the flux deficit. Detailed analysis of the results using the spot modeling for these two stars are described in \cite[Estrela \& Valio (2016)]{estrela16}.

\section{Transit residuals excess}

In the second method of this work we subtracted a modelled light curve of a star without spots from the transit light curves. The result from this subtraction is the residual that clearly shows the spots signatures. An example of this method is shown in Figure \ref{fig:figura1_1} for the 100th transit of Kepler-17. The excess in the residuals (in bold in Fig. \ref{fig:figura1_1}d) corresponds to the spots signatures. Thus, we integrated the residual excess constrained to $\pm$ 70$^{\circ}$ longitude of the star (delimited by the vertical dashed lines of Fig.\ref{fig:figura1_1}). This allows us to characterize the magnetic activity level of the star. To remove any possible trends in this time series, we applied a quadratic polynomial fit and subtracted.

\begin{figure*}[!h]
  \centering
\includegraphics[scale=0.40]{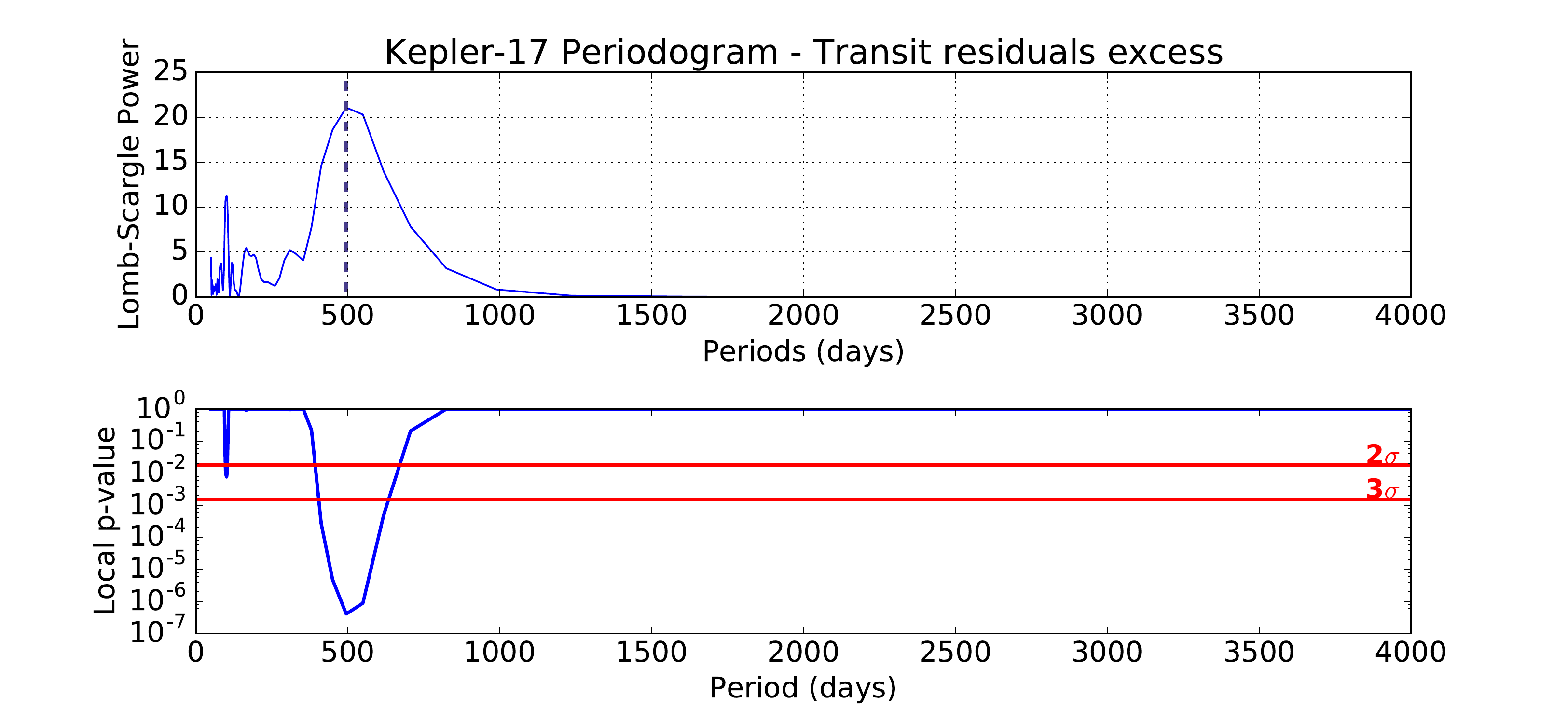}
\includegraphics[scale=0.40]{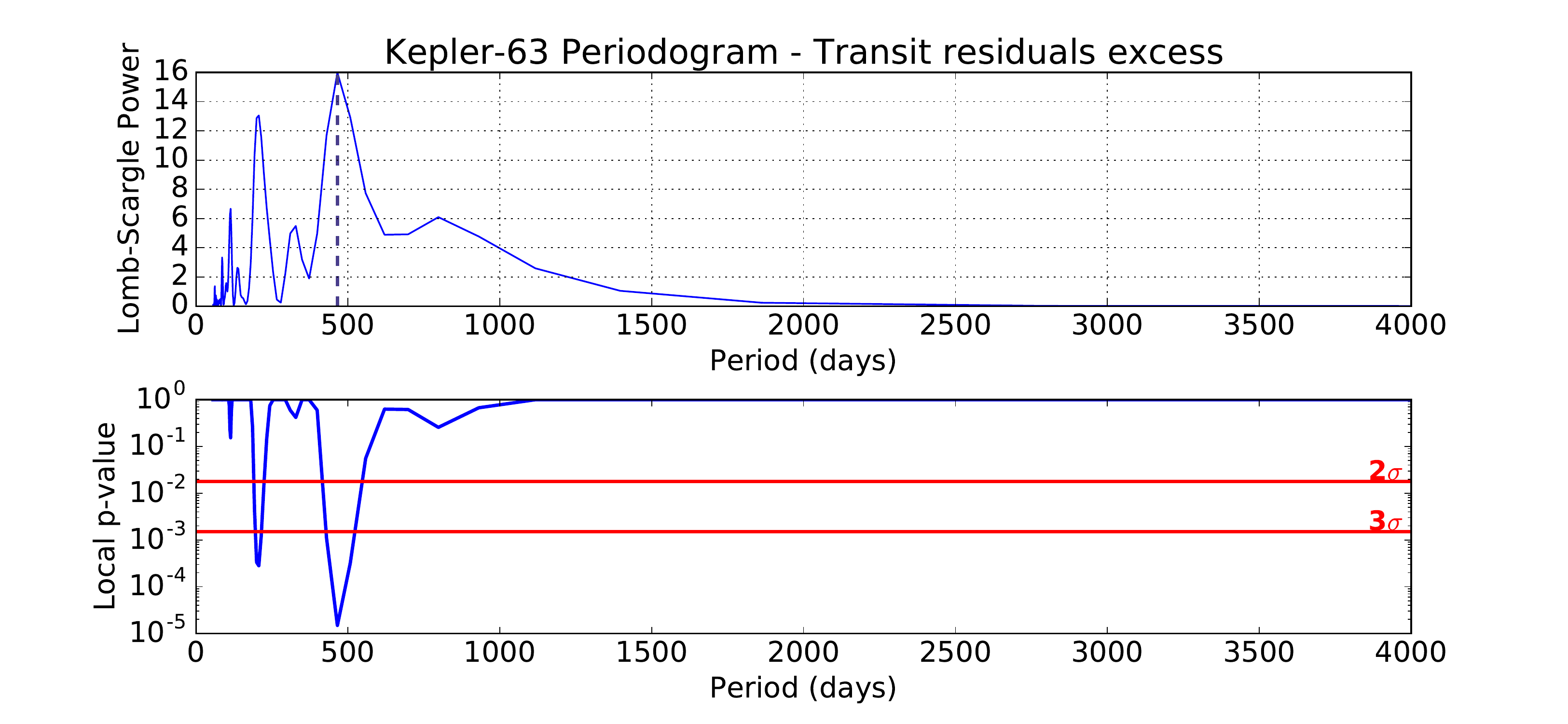}
  \caption[Lomb]{Lomb Scargle periodogram applied to the integrated transit residuals excess of Kepler-17 (\textit{top}) and Kepler-63 (\textit{bottom}). The highest peak, indicated by a dashed line, corresponds to a periodicity of 494 $\pm$ 100 days for Kepler-17 and 465 $\pm$ 40 days for Kepler-63.}
\label{fig:figura1_3}
\end{figure*}

We applied the LS periodogram to the time series resulted from the integrated transit residuals excess. A peak at 490 $\pm$ 100 days was found for Kepler-17 and 460 $\pm$ 40 days for Kepler-63. The value obtained for Kepler-63 is similar to that from the first approach, and corresponds to a 1.27 year-cycle.  On the other hand, the cycle period estimate for Kepler-17 agrees within the uncertainty of the result from the first method. 

Therefore, the results of both methods agree with each other. For this reason, we decided to work with more stars applying only the second method (transit residuals excess), which is easier in estimating magnetic cycles. 

This method was applied to four more Kepler active stars hosting planets: Hat-p-11 (Kepler-3), Kepler-96, KIC 9705459 and KIC 5376836 (see details in Table \ref{table:tab1}). However, the planets candidates orbiting KIC 9705459 and KIC 5376836 were later classified as false positive, and the two stars are now known to be a binary eclipsing system. In this case, the transit of a companion star can also occult a spot in the stellar disk and show its signature, which works similar when eclipsed by a planetary transit. Taking that into consideration, we decided to keep these stars in our analysis. Then, by applying a Lomb Scargle periodogram, we found a clean peak at 305 $\pm$ 60 days for Hat-p-11. Kepler-96 and KIC 9705459 also showed a clean peak at 545 $\pm$ 128 days and 100 $\pm$ 9 days, respectively. Finally, KIC 5376836 showed a significant peak at 42 $\pm$ 3 days. These periodicities show a p-value below the 3$\sigma$ significance level, confirming their significance. Table \ref{table:tab2} shows a summary of the results found for the magnetic cycles of all stars analysed in this work.

\begin{table}[H]
\renewcommand{\arraystretch}{2}
\centering
\caption{}
\label{table:tab2}
\begin{tabular}{lcc}
\multicolumn{3}{c}{\textbf{Magnetic activity cycle periods}}                                                   \\
\hline
Star                & \multicolumn{1}{l}{P$_{\rm cycle}$ (days)} & \multicolumn{1}{l}{P$_{\rm cycle}$ (years)} \\
Kepler-17           & 490 $\pm$ 100                              & 1.35 $\pm$ 0.27 yr                          \\
Kepler-63           & 460 $\pm$ 60                               & 1.27 $\pm$ 0.16 yr                           \\
Hat-p-11 (Kepler-3) & 305 $\pm$ 60                               & 0.83 $\pm$ 0.16                             \\
Kepler-96           & 545 $\pm$ 128                              & 1.50 $\pm$ 0.35                             \\
KIC 9705459         & 100 $\pm$ 9                                & 0.27 $\pm$ 0.024                            \\
KIC 5376836         & 42 $\pm$ 3                                 & 0.11 $\pm$ 0.007         
          
\end{tabular}
\end{table}

\begin{figure*}[!h]
  \centering
\includegraphics[scale=0.40]{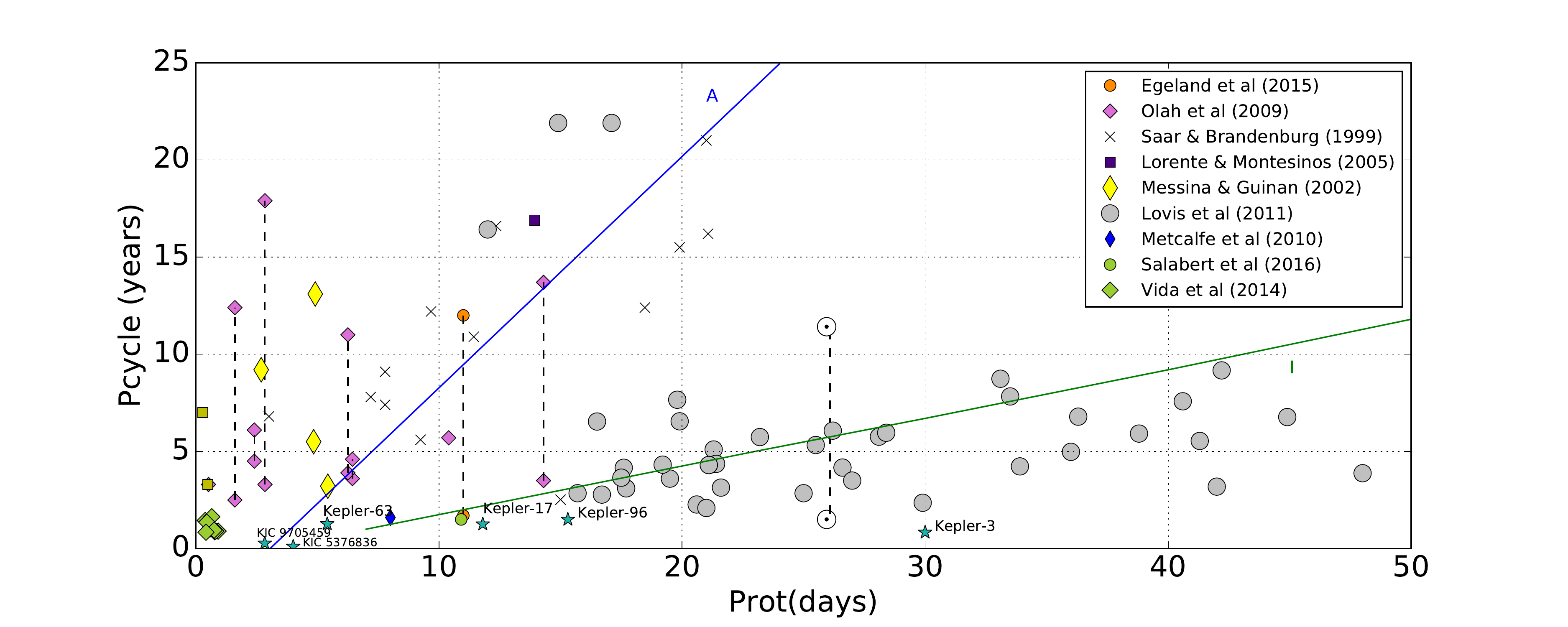}
  \caption[]{Relation between the stellar magnetic cycle, P$_{\rm cycle}$, and rotational period, P$_{\rm rot}$, for a sample of stars from the literature and for the Kepler stars analysed here. The vertical dashed lines connect different cycle periods obtained for the same star (ie.: stars with a short and long cycle). The blue line corresponds to the active branch and the green line is associated with the inactive branch. The star symbol indicates the stars that have the magnetic cycle determined in this work.}
\label{fig:figura1_4}
\end{figure*}

\section{Summary and Conclusions}

We have estimated the period of the magnetic cycle, P$_{\rm cycle}$, for two active solar-type stars, Kepler-17 and Kepler-63, by applying two new methods: spot modelling and transit residuals excess. Since the results of both methods agreed with each other, we used the second method to estimate the magnetic cycle of four more active stars observed by Kepler: Hat-p-11 (Kepler-3), Kepler-96, KIC 9705459 and KIC 5376836. This method is much faster to determine magnetic cycles because it only requires to integrate the area of the residuals due to the activity (spots) in the transit light curve. The first two stars have a transiting planet, while KIC 9705459 and KIC 5376836 had planet candidates, found later to be false positives. These stars were classified as eclipsing binary systems, and we used the primary transit in our analysis to detect spots.

The results found here, with an exception of KIC 5376836 and KIC 9705459, have a P$_{\rm cycle}$ within the same range of 300-900 days found for 9 Kepler fast-rotators stars analyzed by \cite[Vida \& Olah (2014)]{Vida2013}. As we are constrained to the duration of observation ($\leq$ 4 years) of the Kepler telescope, it is not possible to determine longer cycles. In the case of KIC 5376836, this star has only 372 days of observation in the short cadence data, which can justify we have obtained a cycle period of 42 days for this star.

Based on the relation between the stellar rotation period, P$_{\rm orb}$ and stellar cycle period, P$_{\rm cycle}$, proposed by \cite[Saar \& Brandenburg (1999)]{saar99}, we can verify if the stars analysed in this work follow the same trend observed for active stars from literature  (see Figure \ref{fig:figura1_4}). In Figure \ref{fig:figura1_4} one can observe that KIC 5376836 and KIC 9705459 fall within the active branch, while Kepler-63 is located between the active and inactive branches. Kepler-96 and Kepler-17, despite being two active stars, they have their short cycles in the inactive branch. However, as proposed by \cite[B{\"o}hm-Vitense (2007)]{bohm07}, stars in the active branch may also shows secondary cycles that fall in the inactive branch. The cycle period found for these stars are very close to that found for the solar analogue HD 30495 \cite[(Egeland et al., 2015)]{egeland15} and also with the short cycle of 1.5 year found by \cite[Salabert et al.(2016)]{salabert16} for the solar analogue KIC 10644253. In particular, HD 30495 also has another longer cycle period of $\sim$ 12 years and is located in the active branch. However, it is not possible to determine if Kepler-96 and Kepler-17 has a longer cycle period as we have only four years of observational data available. Finally, Hat-p-11 is a special case compared to the other stars in our sample. This star has a long rotation period (P$_{\rm rot}$ $\sim$ 30 days) and has an age older than the Sun (6.5 Gyrs), but it shows remarkable activity. Probably, Hat-p-11 is interacting magnetically with its close orbit Hot-Jupiter, and for this reason its activity level is higher. The short cycle found for this star falls close to the inactive branch. 

It is also important to note that the insertion of the other stars from literature with P$_{cycle}$ recently determined resulted in a spread of the active/inactive branches compared with previous works from \cite[Saar \& Brandenburg (1999)]{saar99} and \cite[B{\"o}hm-Vitense (2007)]{bohm07}.

 \end{document}